# Sustained optothermal propulsion at the air-water interface by pulsating thermocapillary instability

*Rana Zibaei, Maria Giulia Delli Santi, Salvatore Castrignano and Pietro Malara**

Consiglio Nazionale delle Ricerche, Istituto Nazionale di Ottica (INO), via Campi Flegrei 34, 80078 Pozzuoli, Naples, Italy.

*email: pietro.malara@ino.cnr.it

**Abstract**

In an axisymmetric meniscus lifted from a free water surface and locally heated through near infrared radiation by an optical fiber, natural buoyancy and geometric constraints confine the hot liquid near the apex. This confinement enhances the thermal gradient along the meniscus surface and, above a critical optical power, drives the system into an unstable convection regime. By analyzing the refractive index oscillations at the meniscus apex and the simultaneous advection of nearby floating particles, we demonstrate that the instability is characterized by periodic thermocapillary pulses with tunable frequency and directionality. By controlling these features, we modulate the hydrodynamic forces generated by the pulsed surface flow and actuate particles at the meniscus interface in sustained oscillations or stable orbital trajectories. With 20-30mW optical power and particles ranging from 0.1 to 0.5 mm, rotational velocities up to 600 rpm are achieved —the highest reported for continuous optothermal actuation. Our results establish a highly-efficient propulsion mechanism at fluid interfaces with potential applications in light-powered micromotors, interfacial transport and soft microrobotics. More broadly, the optical fiber–meniscus configuration provides a simple and adaptable platform for generating, controlling and detecting large thermocapillary instabilities on fluid surfaces.

## Introduction

Thermally induced surface-tension gradients at fluid interfaces generate Marangoni flows that efficiently convert local temperature differences into mechanical motion. These flows can be easily produced by laser illumination and used to propel floating objects with high responsiveness and minimal power, enabling a growing class of optothermal micromanipulation strategies at the air–water interface and light-driven micromachines [1-4].

Optothermal particle actuation, even in complex trajectories, can be realized by dynamically reconfiguring the illumination pattern on the liquid surface [5-8]. However, achieving sustained motion under steady illumination, which is particularly advantageous for the efficient conversion of light into mechanical work, remains challenging. Persistent oscillations and closed-loop motions of light-absorbing microparticles on the water surface have been observed in laser tweezers setups, arising from the competition between attractive optical forces and repulsive thermal or diffusiophoretic effects [9-11]. Similar dynamics have also been reported on the curved surface of irradiated liquid droplets, where the steady Marangoni propulsion of light-absorbing microparticles is contrasted by capillary restoring forces [12-14]. Sustained rotational motion at a flat air-water interface under steady illumination was also obtained using Janus particles or structures with an asymmetric thermal conductivity. Upon illumination, these structures transfer heat unevenly to the surrounding fluid, generating asymmetric Marangoni flows that result in a net torque [15-17].

In this work, we demonstrate a distinct mechanism capable of propelling particles at the air water interface in sustained, high-velocity oscillatory and orbital trajectories. The described is mechanism directly exploits the hydrodynamic force generated by a thermocapillary pulsed instability in an optically-heated axisymmetric water meniscus.

In a meniscus lifted above a free water surface and irradiated by a vertical optical fiber, buoyancy and geometrical confinement create a highly-localized hot liquid region at the apex. Similarly to other axysimmetric local heating geometries, the thermal gradient on the meniscus surface generates a toroidal convection cell, with an outward radial flow at the surface and an inward replenishment beneath [18-25]. As the meniscus is lifted, the thermal gradient intensifies and these flows become progressively stronger, until they loose stability and transition to an oscillatory state. The convection instability manifests as periodic thermocapillary pulses that originate at the apex and propagate on the liquid surface, exerting a strong advective force on sub-mm particles at the air-water interface.

In the next sections we describe the thermal gradient intensification in the lifted meniscus, characterize the convection instability and harness it to generate sustained-high velocity oscillatory and orbital motions in particles at the air-water interface.

**Experimental setup and steady convection regime**

A single-mode optical fiber (SMF28, 125 $\mu$m diameter) cleaved at right angle is lifted vertically from a pool of ultrapure water (Milli-Q, Millipore) at ambient temperature, in a circular glass container (4 cm diameter, 1 cm depth), producing an axisymmetric liquid meniscus (Figure 1a). The meniscus height $h$ (i.e. the distance between the fiber end and the free water surface), is controlled by a micrometric translation stage. Besides supporting the meniscus, the optical fiber acts as a microscopic heater, by delivering near infrared radiation (NIR) at 1.55 µm from an amplified spontaneous emission source (ASE), and probes the temperature-dependent refractive index of the meniscus water. The fiber interfacial reflectivity is in fact given by the Fresnel relation $R = \left(\frac{n_f - n_w}{n_f + n_w}\right)^2$, with $n_w$ and $n_f$ the refractive indexes of the water and the optical fiber. Since the latter has a fixed value $n_f = 1.45$, the variations of the meniscus backreflected intensity, recorded through a detector behind a 50% fiber splitter, instantaneously map the variations of $n_w$.

In a first set of experiments the meniscus is injected with a NIR power <10mW. The optical fiber, initially in contact with the free liquid surface ($h$=0), is lifted vertically in 50-µm steps while recording the backreflected intensity. As shown in Figure 1b (upper panel) the detector signal increases at every lifting step, indicating a progressive reduction of the water refractive index with $h$.

Experiments with different radiation sources showed that this decreasing $n_w$ trend manifests only upon irradiation with NIR wavelengths strongly absorbed by liquid water [26] and is thus associated to a progressive heating of the meniscus apex during the lifting process. The variations of $n_w$ calculated from the backreflected intensity through the Fresnel relation can be therefore converted in temperature variations $\Delta T(h) = \frac{\Delta n_w(h)}{\frac{dn_w}{dT}}$ using the thermooptic coefficient $\frac{dn_w}{dT} = -1.2 * 10^{-4}$ [27]. The result is shown in fig.1b: when $h$ increases, the meniscus apical temperature increases too, reaching $\Delta T$ ~6K at the maximum equilibrium height.

This temperature increase, occurring at constant energy input just by lifting the meniscus, can be intuitively attributed to the buoyancy-driven uplift of hot water and the lateral capillary confinement. As illustrated in the sketch of Figure 1c (upper panel), in the lifted meniscus hot water is pushed towards the radiation source and confined at the apex, where the capability to exchange heat with the colder reservoir liquid is strongly limited.

As the apical water heats up, the thermal gradient along the meniscus surface increases and thermocapillary transport becomes an increasingly important pathway to dissipate the radiation-absorbed heat. This mechanism reflects in the increase of the system's Marangoni number vs *h*, plotted in fig.1c (bottom panel). The $\Delta M_a(h)$ values were calculated as $\Delta M_a(h) = \frac{\partial\gamma}{\partial T}\frac{\Delta T(h) * L(h)}{\mu\alpha}$ using the experimental points $\Delta T(h)$ of Figure 1b, a characteristic length $L(h) \approx 2h, \mu = 8.9 \cdot 10^{-4}$ Pa/s and $\alpha = 1.4 \cdot 10^{-7}$ $m^2/s$ (water dynamic viscosity and thermal diffusivity respectively).

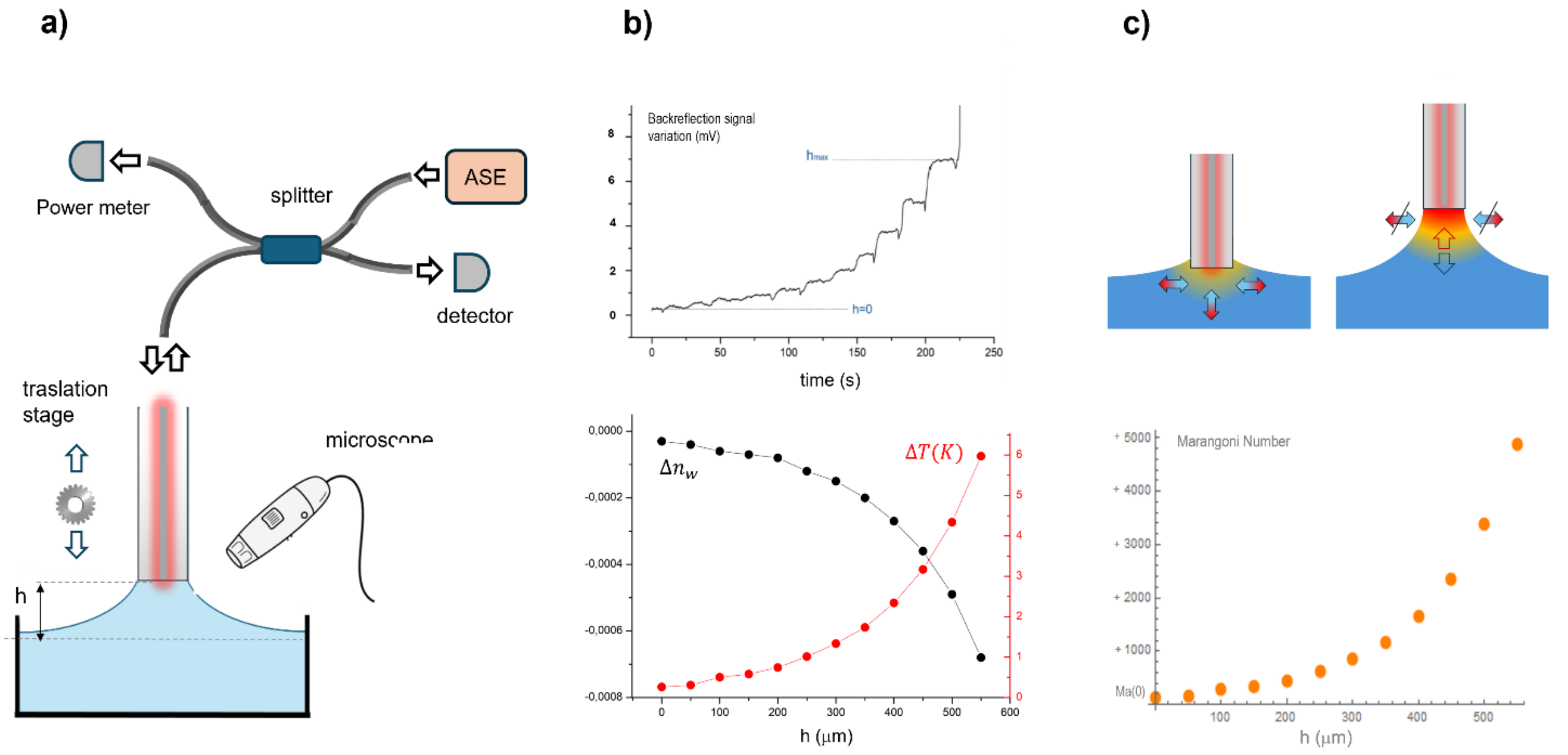

***Figure 1. Temperature increase of a lifted water meniscus**. a) experimental setup: a right-angle cleaved optical fiber lifts an axysimmetric water meniscus from a free surface. The fiber heats the meniscus through near infrared radiation and collects the intensity backreflected by the fiber-water interface; b) top panel: example of fiber backreflection signal recorded with 10mW optical power while lifting the meniscus in 50-$\mu m$ steps until it breaks. Bottom panel: refractive index and meniscus temperature values calculated from the backreflected intensity; c) top panel: illustration of the meniscus heat dissipation: when lifted, buoyancy and meniscus walls confine hot water near the apex. The process inhibits the bulk heat-exchange channels in favor of surface thermocapillary transport, increasing the system's Marangoni number (bottom panel).*

**Instability convection regime**. A second set of experiments was conducted with a larger NIR input power, in the 15-30mW range. In these experiments, an oscillatory temperature instability regularly manifests when the meniscus is lifted. Figure 2a shows an example of fiber backreflection signal (which positively correlates with the apical water temperature) recorded with @20 mW input optical power while varying the meniscus height $h$ as illustrated in the bottom panel.

- (0-25s): the backreflection initially reproduces the same increasing-temperature pattern of Figure 1b, but transitions to a pulsed behavior when the critical meniscus height $h_1 \sim 350\ \mu m$ is passed.
- (25-50s): by further increasing $h$, the pulse frequency increases and the signal evolves towards a triangular waveform (see Figure 2b). At the same time, the oscillation amplitude and the mean signal level decrease, indicating that in this oscillatory regime the meniscus cools down while lifted, as opposed to the low-power experiments of fig.1.
- (50-60s): when a second threshold elevation is passed ($h_2 \sim 380\ \mu m$) the water temperature becomes steady again, and approximately equal to the $h$=0 condition.
- (60-90s) when the meniscus is lowered again below $h_2$, the meniscus temperature reproduces backwards the same oscillatory patterns of the lifting process and consistently transitions to the steady temperature phase when also $h_2$ is crossed in the opposite direction. In between $h_2$ and $h_2$ the oscillation features are completely reproducible and therefore controllable.

The oscillatory dynamics of the backreflection signal reflect a periodic temperature build-up at the meniscus apex followed by a sudden release and recovery. At low $h$ values, the slow periodicity of these cycles ensures a significant time for thermal recovery, resulting in large temperature pulses. At higher $h$ values, when the cycles get faster, there is less time for heat accumulation and the pulses amplitude consequently reduces. Importantly, the average meniscus temperature also reduces (mean signal value decreases), showing that, in this oscillatory regime, lifting the meniscus improves heat dissipation, thus reducing the apical temperature. In fact, the pulsation extinguishes after a critical height $h_2$.

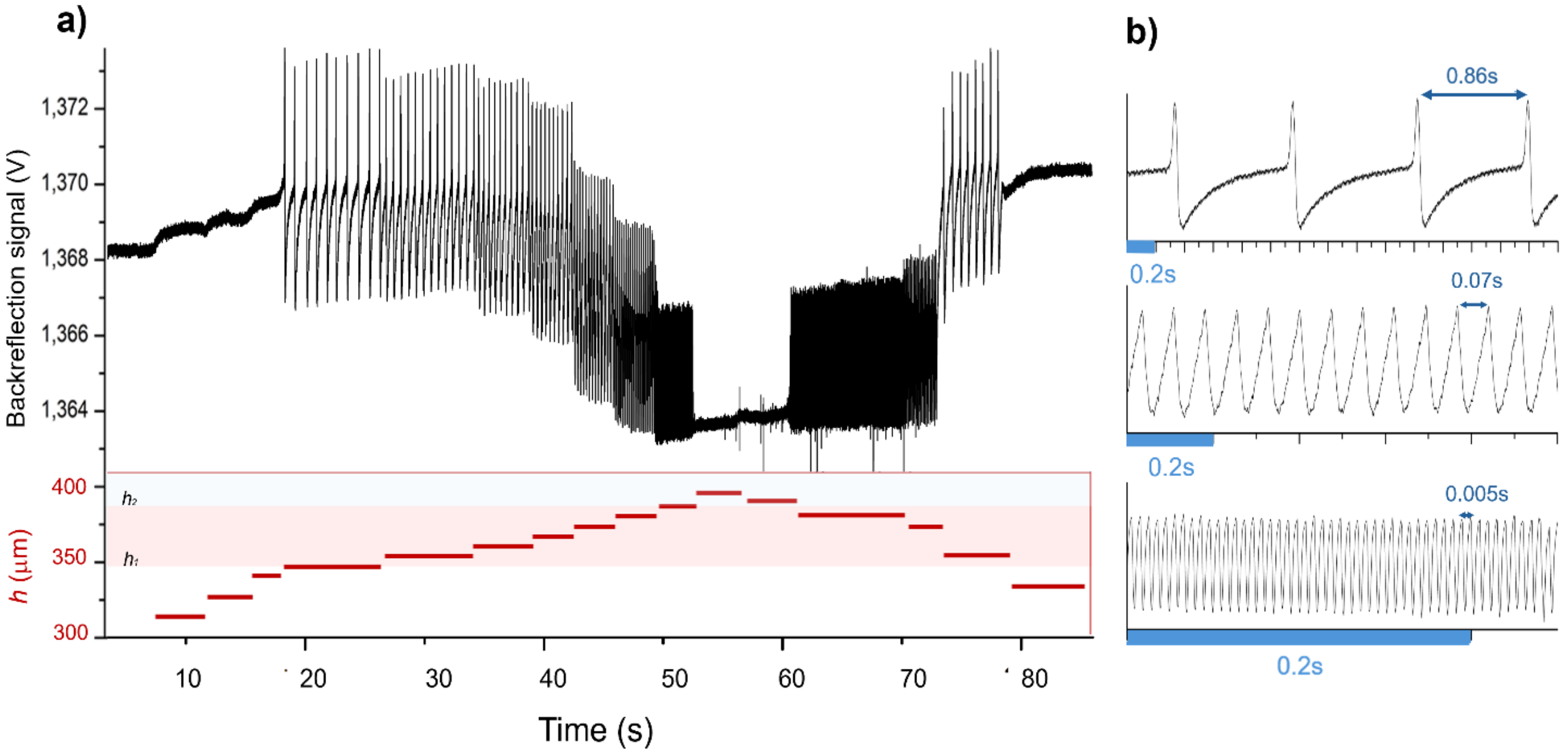

***Figure 2. Convection instability in a lifted water meniscus****. a) Upper panel: oscillatory dynamics of the backreflection signal with input NIR power >15mW. The signal positively correlates with the apical water temperature. Bottom panel: corresponding meniscus height over time. b) normalized backreflection signal for 3 increasing h values, highlighting the increase of the oscillation frequency.*

Additional insight in the nature of the described meniscus instability can be gained through the observation of particles at the water surface. At every temperature pulse, floating particles within a few mm from the meniscus axis are advected with an extremely rapid forward–backward radial motion, with peak velocities in the order of ~1cm/s. This is shown in the attached clip vid01, recorded with a 30f/s digital microscope placed underneath the glass water container. A frame-by-frame analysis of this video, carried out in the supplementary materials, reveals that:

- The perturbation responsible for the particles advection originates at the meniscus apex and propagates outwards on the water surface. Indeed, the delay between the detection of a pulse and the advection of the particles increases with their radial distance from the meniscus axis. By measuring this delay, the propagation velocity $v_p$ of the advective perturbation can be estimated. We selected from vid01 advection events involving multiple particles at the same time, and plotted their axial coordinates against the onset times of their motion. Since the onset of motion of a particle marks the instant at which the perturbation reaches it, $v_p$ is given by the slope of a linear fit to these data. An example is in fig.3a, relative to an advection event with 4 particles arranged as in fig.3b. This analysis, repeated over several multiparticle events, returns an average perturbation velocity $v_p = 21 \pm 1\ mm/s$ (see also supplementary materials).

- The rapid forward-backwards trajectory of the advected particles suggests that the perturbation is a finite-width, high-temperature surface pulse. A finite-width hot water front carries two temperature gradients on the water surface, with respect to the colder liquid ahead-of and behind it, thereby producing two counter-directed thermocapillary flows at its leading and trailing edges. As illustrated in Figure 3c, when the high-temperature front first approaches a particle, its leading flow pushes it forward with an initial velocity $v_f$. However, if $v_f < v_p$, the hot front reaches and then surpasses the advected particle, exposing it to the trailing thermocapillary flow, which abruptly inverts its motion. Consistently with this picture, in all our observations $v_f \simeq 0.5 v_p$ and the backwards velocity $v_b$ is always smaller than $v_f$ (see supplementary materials).

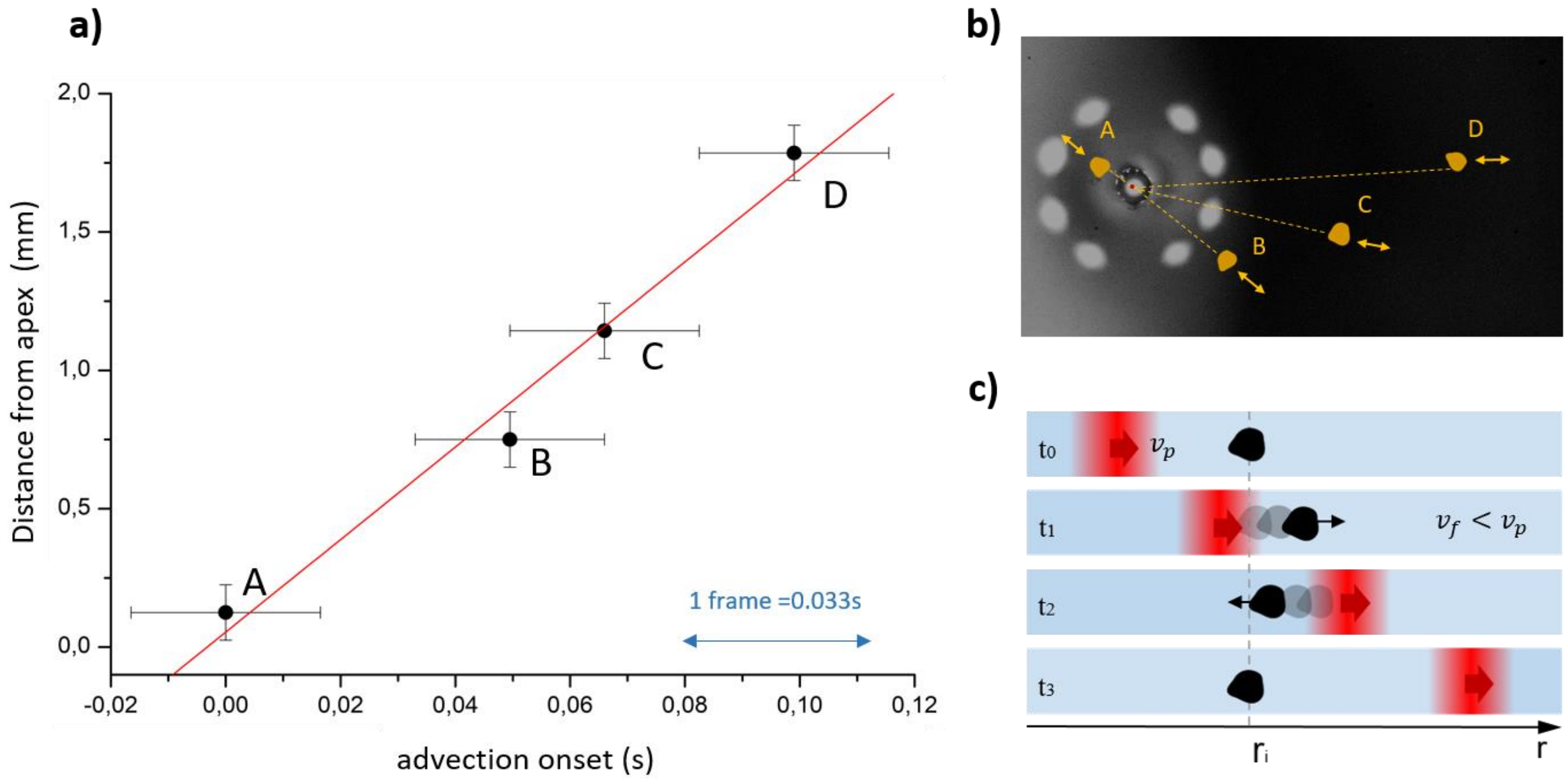

***Figure 3. Advection of floating particles near the unstable meniscus****. a) positions of 4 particles A, B, C, D against their advection onset times after a single temperature pulse. The propagation velocity $v_p$ of the advecting perturbation is estimated through a linear fit of this data. Repeating the analysis for multiple pulse events yields an average value $v_p = 2.1 \pm 0.1\ cm/s$. b) initial positions of the advected particles in the microscope field of view (view from below: the crown of*

*white spots around the fiber facet is the reflection of the microscope lights on the water surface); c) the rapid forward–backward motion of the particles is consistent with the passage of a localized hot thermal pulse. As the pulse approaches, its thermal gradient drives the particles forward. However, since the pulse propagates faster than the particles themselves, it rapidly overtakes them, exposing them to a thermal gradient of opposite sign and causing the abrupt reversal of their motion.*

The process emerging from our experimental observations, periodic heat accumulation and release trough the emission of thermocapillary pulses, bears some similarities to the pulsating regimes and oscillatory, target-like temperature patterns reported for other axisymmetric fluid-heating configurations [19, 21, 22, 28]. However, in none of these cases the pulsation is associated to a finite-width thermal front propagating away from the heat source. For a full characterization of the meniscus instability further investigations are necessary, particularly through thermal imaging of the liquid surface and tracer-particle mapping of the bulk flow.

It is finally worth noting that both the temperature dynamics of fig.2 and the particles advection of vid01 manifest consistently when the instability onsets. However, the specific threshold values $h_1$, $h_2$ and the NIR power required to initiate the instability depend significantly on the water sample used (and can change over a few hours even with the same sample). This is likely due to the unavoidable contamination of the water surface, i.e. the progressive formation of a thin adsorption layer that influences the stability and the structure of the convection flow induced by heat [9,28].

**Sustained propulsion of particles in a capillary trap.**

The impulsive force that advects the floating particles in the unstable regime can be harnessed to generate a sustained motion if a restoring mechanism returns the propelled body to its initial position between successive impulses. For this task we employed irregular fragments of polypropylene (PP) and acrylic urethane (AU), with sizes ranging from 10 to 500 μm. The lateral capillary interaction of these particles with the fiber meniscus is attractive [29, 30], and dominates at distances shorter than the water capillary length. On the other hand, since PP and AU are buoyant and therefore bound at the water surface, the component of gravity tangential to the meniscus surface generates a net force directed away from the fiber axis. This repulsion dominates near the fiber edge, where the local slope of the meniscus is larger. Together, capillary attraction and gravity repulsion establish a potential well that traps the PP and AU fragments at an equilibrium position $r_{eq}$, corresponding to a local potential energy minimum (Figure 4a). The specific value of $r_{eq}$ depends on both the particle shape and mass as well as on the meniscus height.

When a thermocapillary flow develops on the meniscus surface, it exerts a hydrodynamic drag force $F_H$ on the trapped particles. For small Reynolds number (as in our case), $F_H \propto u - v$, where $u$ is the surface flow velocity at the particle's position (proportional to the local temperature-driven surface tension gradient $\frac{d\gamma}{dT}\nabla T$) and $v$ is the velocity of the particle itself.

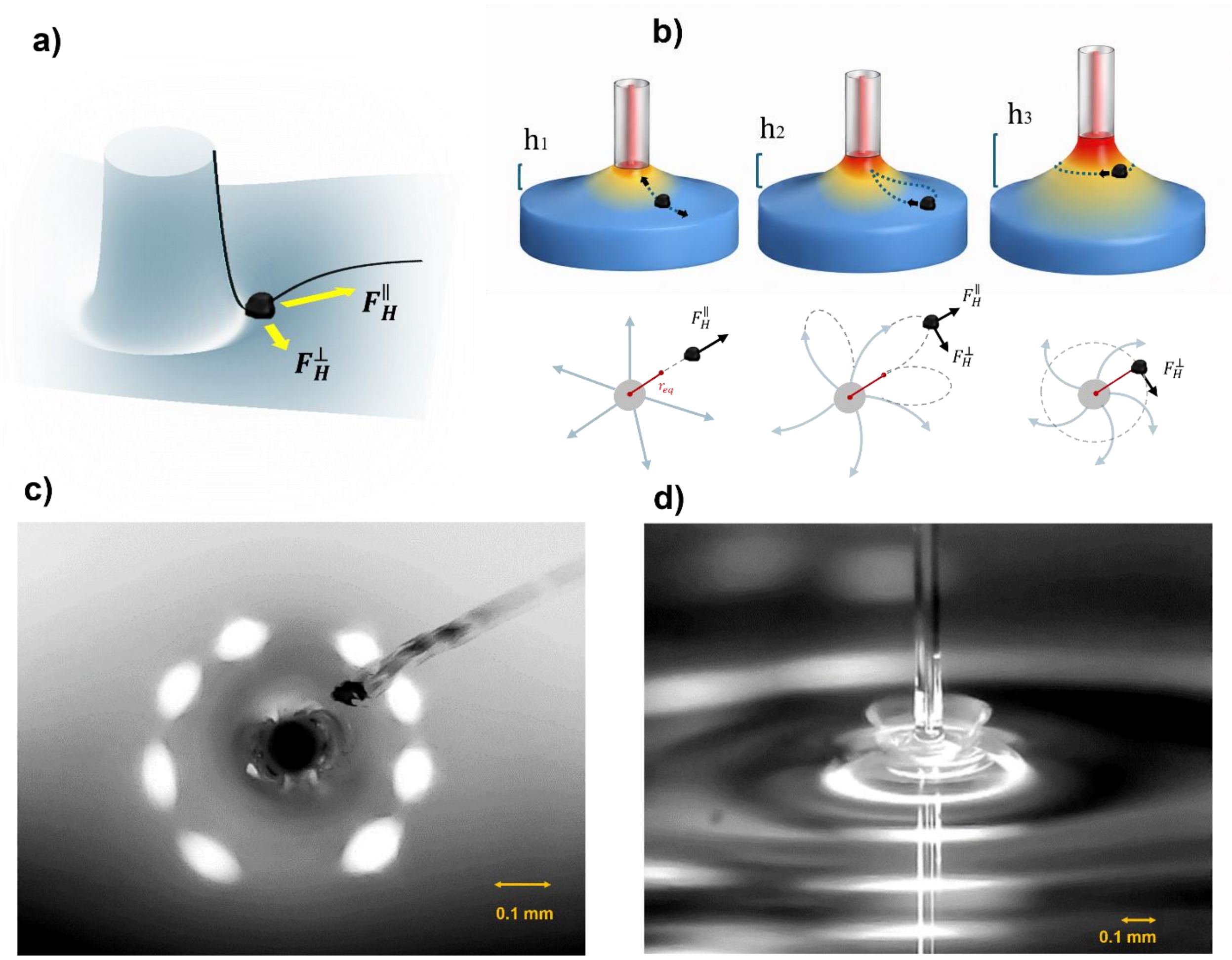

***Figure 4. Sustained propulsion of capillary-trapped particles**. a) potential energy landscape established near the meniscus by gravity and lateral capillary forces, and hydrodynamic force*

*components $F_H^{\parallel}$, $F_H^{\perp}$ acting on a trapped particle. b) particle trajectories observed for three increasing values of the meniscus elevation and hydrodynamic force field configurations inferred from them (blue arrows). For small h values the thermocapillary flow is radial, so only the component $F_H^{\parallel}$ exists and drives radial oscillations. For increasingly larger h the thermocapillary flow spiralizes and the azimuthal component $F_H^{\perp}$ emerges; the fragments trajectories first evolve to petal-like patterns, then transition to a circular orbit with radius $r_{eq}$; c,d) experimental observations of oscillating and orbiting particles: frames overlay from vid02 and vid04 respectively.*

Under all steady conditions, the particles remain still at their equilibrium position. However, as soon as the instability onsets, the hydrodynamic drag of the surface convection pulses generates different types of periodic motions, as illustrated in fig.4b.

<u>*Radial oscillations*</u>. In the early stages of the instability (low *h* values) particles are rapidly propelled outwards and, in the time between pulses, drawn back at their equilibrium position. This continuous oscillatory motion can be tuned in real time up to several Hz by tuning *h*, i.e. the pulsation rate. Examples can be found in vid02 and vid03 (PP and AU fragments respectively), and a typical trajectory is also visualized in Figure 4c with a frame overlay from vid02.

<u>*Petal-like trajectories*</u>. When the meniscus is slightly lifted, the above described radial oscillations tend to evolve into petal-like trajectories (observable at the end of vid02), which reflect the emergence of an azimuthal component of the pulses hydrodynamic force ($F_H^{\perp}$) alongside the radial one ($F_H^{\parallel}$). The appearance of the component $F_H^{\perp}$ and its growth with increasing *h* is consistent with the breakup of the radial symmetry of the surface flow into an azimuthal spiral pattern, an effect widely documented in fluids under large axisymmetric temperature gradients [18-25].

<u>*Orbital motions*</u>. By further increasing the meniscus elevation, the particle oscillations transition into a continuous orbital motion around the meniscus. Experimental observations of such motions with different particles are reported in vid04 (in sequence: PP fragment, AU fragment, large AU raft), and a typical trajectory is visualized in fig.4d with a superposition of frames from the same video; clear transitions from the oscillatory to the orbital trajectory upon meniscus lifting can be observed in vid05 and vid06. The particles orbital velocity depends largely on their size and geometry, but it can be controlled through the meniscus height and, to a lesser extent, with the NIR radiation power. This is clearly shown in vid07, where, by tuning these parameters, an unidentified, elongated particle trapped in the meniscus potential well is progressively accelerated up to 600 rpm, to our knowledge the largest velocity achieved for sustained optothermal actuation.

The transition from oscillatory to orbital trajectory is consistent with the instability pulsation trends observed in fig.2. As the meniscus is lifted, the pulses amplitude progressively reduces, necessarily leading to a point where $F_H^{\parallel}$ can't displace anymore the particle from the bottom of its radial potential well. In this situation, the uncontrasted, growing azimuthal component $F_H^{\perp}$ becomes the only force capable of actuation. At the same time, the increasing oscillation frequency becomes too fast for the particles dynamics to follow, and they start responding to a time-averaged hydrodynamic force, resulting in a smooth and continuous circular motion.

## Conclusions

In an axisymmetric water meniscus irradiated from above with near infrared radiation, buoyancy and geometric confinement concentrate thermal energy near the apex, amplifying the thermocapillary stress and enabling continuous tuning of the Marangoni number across a wide range. For increasing confinement, we observe a first transition to an oscillatory convection regime, followed by a second steady temperature state. We experimentally characterized the oscillatory convection phase, demonstrating that the thermocapillary flow can be shaped into periodic pulses with programmable frequency and a velocity field that can be tuned from a radial to a spiral pattern. Controlling these features allows to modulate the hydrodynamic force exerted by the thermocapillary pulses on particles at the air–water interface and actuate them. In particular, by exploiting a gravity-capillary trap, we employed the pulsing instability to drive sustained oscillating or orbital motion of macroparticles trapped in the meniscus vicinity. Unlike other cyclic motions reported with steady Marangoni flows, this propulsion effect does not rely on a secondary heat dissipation of irradiated particles on the liquid surface. As such, it does not directly heat the actuated bodies, requires no special particle shape, and can actuate particles from the microscale to the submillimeter scale. The achieved velocities, reaching the centimeter-per-second range, are the fastest reported for persistent optothermally-induced motions, representing an efficient conversion of light into mechanical work with the potential to enable new applications in soft microrobotics and light-powered micromotors.

**Supplementary material**

Further details on the particle advection experiments, providing evidence that heat is released by the meniscus as periodic, finite-width, high-temperature pulses originating at the apex and traveling outward along the liquid surface.

**Associated content**

The following multimedia files are also associated to the manuscript:

- vid01.mp4: forward-backwards advection pattern of free-floating particles during the pulsating instability.vid02.mp4: continuous oscillation of a ~100 μm Acrylic Urethane fragment (view from below)
- vid03.mp4 : continuous, fast radial oscillation of a ~10 μm polypropylene fragment (top view)
- vid04.mp4: 3 examples of continuous orbital motion (PP fragment, AU fragment, large AU raft)
- vide05.mp4 oscillatory-to-orbital transition of a polypropylene fragment (top view)
- vid06.mp4 oscillatory-to-orbital transition of large Acrylic Urethane fragments (view from below)
- vid07.mp4: acceleration of an orbiting fragment up to 600 rpm by tuning the meniscus height and the input power between 20 and 30 mW (unidentified fragment)

**Funding Sources**

This research was supported by the Italian Ministry of University and Research within the PRIN 2022 program, Project ID-SWAP "interferometric detection of stretched water's properties and their role in biological systems". 2022YZEXBL_PE2_PRIN2022